\documentclass[sigconf]{acmart}

\usepackage[utf8]{inputenc}
\usepackage[T1]{fontenc} 

\usepackage{natbib}

\usepackage{algorithmic}
\usepackage{algorithm}
\usepackage{array}
\usepackage[caption=false,font=normalsize,labelfont=sf,textfont=sf]{subfig}
\usepackage{textcomp}
\usepackage{stfloats}
\usepackage{xurl}
\usepackage{verbatim}
\usepackage{graphicx}
\usepackage{tikz}
\usepackage{fontawesome5}
\usetikzlibrary{arrows.meta,calc,fit,backgrounds,matrix}

\DeclareRobustCommand{\rqIcon}[1]{\makebox[2.6mm][c]{\raisebox{\dimexpr(\fontcharht\font`X-1.9mm)/2+\depth\relax}{\resizebox*{!}{1.9mm}{#1}}}}
\DeclareRobustCommand{\rqVaried}{\rqIcon{\faSlidersH}}
\DeclareRobustCommand{\rqFixed}{\rqIcon{\faLock}}

\usepackage{hyperref}
\usepackage{csquotes}
\usepackage{enumitem}

\usepackage{colortbl}
\usepackage{xcolor}
\usepackage{multirow}
\usepackage{tabularx}
\usepackage{booktabs}
\usepackage[super]{nth}
\usepackage{balance} 

\newcommand*{\enq}[1]{\enquote{{\itshape#1}}}

\copyrightyear{2026}
\acmYear{2026}
\setcopyright{cc}
\setcctype{by}
\acmConference[JAWs@ASE 2026]{2nd Journal Ahead Workshop}{October 12--16, 2026}{Munich, Germany}
\acmBooktitle{2nd Journal Ahead Workshop (JAWs@ASE 2026), October 12--16, 2026, Munich, Germany}
\acmDOI{}
\acmISBN{}

\begin{document}
\sloppy

\title{Loop Engineering: Building Blocks, Adoption, and Impact}

\author{Jai Lal Lulla}
\affiliation{%
  \institution{Singapore Management University}
  \city{Singapore}
  \country{Singapore}}
\email{jailal.l.2025@phdcs.smu.edu.sg}

\author{Vahram Nersesyan}
\affiliation{%
  \institution{Heidelberg University}
  \city{Heidelberg}
  \country{Germany}}
\email{vahram.nersesyan@stud.uni-heidelberg.de}

\author{Seyedmoein Mohsenimofidi}
\affiliation{%
  \institution{Heidelberg University}
  \city{Heidelberg}
  \country{Germany}}
\email{s.mohsenimofidi@uni-heidelberg.de}

\author{Christoph Treude}
\affiliation{%
  \institution{Singapore Management University}
  \city{Singapore}
  \country{Singapore}
}
\email{ctreude@smu.edu.sg}
\orcid{0000-0002-6919-2149}

\author{Sebastian Baltes}
\affiliation{%
  \institution{Heidelberg University}
  \city{Heidelberg}
  \country{Germany}
}
\email{sebastian.baltes@uni-heidelberg.de}
\orcid{0000-0002-2442-7522}

\renewcommand{\shortauthors}{Lulla et al.}

\begin{abstract}
Over the past months, the way developers guide agentic AI coding tools has moved up several levels of abstraction, from phrasing prompts to engineering context to configuring the harness around the underlying model. In June 2026, practitioners began to describe a new level called \emph{loop engineering}: Instead of prompting an agent interactively, developers design systems that prompt agents for them. The term spread rapidly, accompanied by bold claims and vocal skepticism, but its adoption in software projects has not yet been measured. We present (1) an exploratory review of the emerging gray literature on loop engineering, (2) a traceability analysis and an exploratory mining study of 36,710 software repositories, and (3) the design of a controlled experiment to study how loop goals, static analysis checks, and decision policies affect task success, code quality, resource usage, and token consumption of agent loops. Our literature review found broad agreement on what a well-engineered loop contains: agent loops started on a schedule or by repository events and bounded by machine-checkable stop conditions, persistent state files, verifier subagents, token budgets, and defined points of escalation to humans. These sources provide no systematic evidence of the claimed benefits. In the mining study, we found autonomous agent loops in 217 open-source GitHub repositories using heuristics and behavioral signals. Nearly all of these repositories committed the configuration that starts their loops, but almost none of them committed the state files that the gray literature suggests. Our proposed experiment is a first step toward closing the evidence gap identified in our literature review. We call for further empirical studies of the benefits and costs of loop engineering for software engineering tasks.
\end{abstract}




\maketitle

\section{Introduction}
\label{sec:introduction}

\looseness=-1 Agentic AI coding tools such as Claude Code or OpenAI Codex interpret a goal, decompose it into steps, invoke tools, and iterate until they consider the goal achieved~\cite{galster2026harness}.
For most developers, working with these tools still entails typing a request, reading the result, and then typing the next request.
However, some prominent practitioners claim to have abandoned this interactive mode altogether.
In an interview published on 2~June 2026, Boris Cherny, the creator of Claude Code~\cite{cherny2026post}, said: \enq{I don't prompt Claude anymore. I have loops that are running. They're the ones that are prompting Claude and kind of figuring out what to do. My job is to write loops}~\cite{cherny2026unplugged}.
On 7~June 2026, Peter Steinberger, the author of OpenClaw, posted a similar message: \enq{you shouldn't be prompting coding agents anymore. You should be designing loops that prompt your agents}~\cite{steinberger2026post}, a post that had reached more than eight million views by mid-July~\cite{orosz2026newsletter}.
The same day, Addy Osmani published a blog post naming the emerging practice \emph{loop engineering} and defining it as \enq{replacing yourself as the person who prompts the agent. You design the system that does it instead}~\cite{osmani2026loop}.

\looseness=-1 The term \emph{loop engineering} extends the trajectory from prompt engineering via context engineering to harness engineering (see Section~\ref{sec:background-trajectory}).
In fact, the practitioner discourse places loop engineering one abstraction level above harness engineering. Where a harness configures an agent loop, a loop controls repeated agent loops over time. It specifies what starts each agent loop (e.g., a schedule or an event such as a failing build or an opened issue), how its results are verified and persisted, and when to escalate to a human~\cite{osmani2026loop, huashu2026playbook}.

\looseness=-1 We argue that this development deserves the attention of empirical software engineering researchers.
First, the practice is no longer hypothetical: Between March and May 2026, tool vendors added commands such as \texttt{/loop} and \texttt{/goal} that developers can use to trigger recurring and goal-driven agent loops~\cite{orosz2026newsletter}.
Second, the claims attached to loop engineering rest almost entirely on anecdotes and self-reported productivity numbers, while experienced engineers call loop engineering \enq{a renamed cron job} and warn about token costs and reviewer fatigue~\cite{orosz2026post, orosz2026newsletter}.
Third, the past years have shown that practitioner vocabulary can influence research agendas quickly. For example, the discourse-driven term \emph{vibe coding} attracted surveys and empirical studies within months~\cite{sapkota2025vibe, ge2025vibecoding}.
Our goal is to move loop engineering from a practitioner term to an empirically grounded object of study, through three research questions:

\begin{description}[leftmargin=1.5em]
\item[RQ1] \emph{How is loop engineering defined in gray literature, and how does it relate to prompt, context, and harness \mbox{engineering?}}
\item[RQ2] \emph{Which aspects of loop engineering are traceable from repository data, and which practices are adopted in open-source projects?}
\item[RQ3] \emph{How do loop goals, static analysis checks, and decision policies affect task success, code quality, resource usage, and token consumption of agent loops?}
\end{description}

\looseness=-1 We answer \textbf{RQ1} by situating loop engineering in the trajectory from prompt via context and harness engineering (Section~\ref{sec:background}) and introducing a working definition (Section~\ref{sec:background-emergence}), then synthesizing selected gray literature into six observations (Section~\ref{sec:review}). We address \textbf{RQ2} with a traceability analysis and an exploratory mining study of 36,710 engineered software repositories (Section~\ref{sec:agenda-mining}). To answer \textbf{RQ3}, we describe a controlled study of loop goals, static analysis checks, and decision policies (Section~\ref{sec:agenda-experiment}) that we will execute for the planned journal extension.

\section{Background}
\label{sec:background}

\subsection{From Prompts to Context to Harnesses}
\label{sec:background-trajectory}

\looseness=-1 \emph{Prompt engineering} was one of the first named practices for guiding large language models (LLMs). Catalogs of reusable prompt patterns~\cite{white2023promptpatterns} and a broad survey of prompting techniques~\cite{schulhoff2024promptreport} systematized how to phrase prompts, and in software projects prompts soon became engineering artifacts in their own right, committed to repositories and evolving like code~\cite{tafreshipour2025prompting, villamizar2025prompts}.
\emph{Context engineering} generalized this practice from single instructions to the systematic design of everything a model sees in its context window~\cite{mei2025context, hua2025context20}.
For agentic coding tools, its most visible instrument is a repository-versioned context file (e.g., \texttt{AGENTS.md} or \texttt{CLAUDE.md}), a \enq{README for agents}~\cite{agentsmd2026}, with mining studies examining how such files are structured and how they evolve~\cite{mohsenimofidi2026context, chatlatanagulchai2025agentreadmes}.
Early evidence on their effect is mixed: One study associated \texttt{AGENTS.md} presence with lower runtime and fewer output tokens~\cite{lulla2026agentsmd}; another found no gain in task success at over 20\% higher inference cost~\cite{gloaguen2026evaluating}.

\looseness=-1 \emph{Harness engineering} moves the focus from the context to the harness, i.e., \enq{the software layers around the model(s) that drive the agent loop}~\cite{galster2026harness, ning2026codeharness}.
The \emph{agent loop} is the model's repeated cycle of calling tools and acting on the results until a stop condition ends it, such as a final response, an achieved goal, or a time limit. \emph{Loop engineering} adds structures that start an agent loop, verify its results, and start it again over time.
For example, an agent can run tests and revise a change within one session, while a loop can start new sessions whenever a build fails.
In an exploratory study of 2{,}853 open-source repositories, \citeauthor{galster2026harness} mapped eight configuration mechanisms (context files, settings, skills, subagents, commands, hooks, rules, and Model Context Protocol (MCP) servers) across five agentic coding tools. They found that developers overwhelmingly adopt static context files and rarely use executable mechanisms, concluding that \enq{harness engineering in open source today is \ldots mostly context engineering}~\cite{galster2026configuring, galster2026harness}.
Their taxonomy contains no mechanism that determines \emph{when and how often} an agent runs.
Hooks are a related mechanism but fire at fixed lifecycle points within a session.
Therefore, loop recurrence, termination, state across agent loops, budgets, and graduated human oversight are all beyond the scope of that analysis.
With our study, we intend to close that gap.

\subsection{The Emergence of Loop Engineering}
\label{sec:background-emergence}

\looseness=-1 Developers ran agents in loops before the term \emph{loop engineering} existed, at two different levels.
The first level is a single execution of an agent loop: In March 2024, Andrew Ng reported that GPT-3.5 \enq{wrapped in an agent loop} reached up to 95.1\% on HumanEval, against 67.0\% for zero-shot GPT-4~\cite{ng2024agentic}.
The second level adds recurrence and state across agent loops, and it arrived as a shell loop. In July 2025, Geoffrey Huntley published the \enq{Ralph Wiggum} technique, \enq{in its purest form, [\ldots] a Bash loop [\ldots] \texttt{while :; do cat PROMPT.md | claude-code ; done}}~\cite{huntley2025ralph}. It works around the finite context window by restarting the agent with a fresh context in every iteration while persisting progress to the file system~\cite{orosz2026newsletter}.
Tool vendors later adopted this technique: Claude Code added a recurring \texttt{/loop} command in March 2026, Codex introduced goal objects with machine-checkable completion conditions in April, and Claude Code, Hermes, and other harnesses followed with \texttt{/goal} commands in May 2026~\cite{orosz2026newsletter}.

\looseness=-1 The term \emph{loop engineering} was coined in early June 2026, when Steinberger's post, Cherny's remark, and Osmani's blog post appeared within days of each other~\cite{steinberger2026post, osmani2026loop, orosz2026newsletter}. Later that month Ng called it \enq{a hot buzzphrase}~\cite{ng2026loops}.
Osmani described a loop as \enq{a recursive goal where you define a purpose and the AI iterates until complete}, with state persisted across agent loops (\enq{the agent forgets, the repo doesn't})~\cite{osmani2026loop}.
A month later, he reported a four-part classification of loops ordered by how much the developer hands off: \emph{turn-based} loops keep each turn under the developer's review, \emph{goal-based} loops hand off the stop condition, \emph{time-based} loops hand off the execution trigger, and \emph{proactive} loops hand off the prompt itself~\cite{osmani2026monthon}.

\looseness=-1 Within weeks, an ecosystem formed around these ideas.
In a community repository, Cobus Greyling distilled the discourse into seven loop patterns (e.g., `daily triage' and `CI sweeper' for continuous integration), four \enq{readiness levels} from documented intent (L0) to unattended operation (L3), catalogs of failure modes and anti-patterns, and tooling for scaffolding, auditing, and cost estimation~\cite{greyling2026repo}.
Andr\'e Lindenberg framed loop engineering as \enq{a different solution architecture, not a different way to use the agent} and argued that the verifier must be designed first~\cite{lindenberg2026loops}.
In a self-published playbook, HuaShu decomposed one loop turn into five \enq{moves} (discovery, handoff, verification, persistence, scheduling)~\cite{huashu2026playbook}. Ng distinguished the agentic coding loop (minutes) from the developer and external feedback loops around it (hours to weeks)~\cite{ng2026loops}.

\looseness=-1 For each of the four layers of guiding AI coding agents, Table~\ref{tab:layers} names the unit of concern and example artifacts. Within the harness layer, \citeauthor{boeckeler2026harness} distinguishes what is built into a coding agent (e.g., the system prompt) from what developers configure on top~\cite{boeckeler2026harness}. We call these the \emph{inner} and \emph{outer harness}, and only artifacts of the outer harness can appear in the repository of a project adopting loop engineering.

\looseness=-1 We have consolidated the practitioner discourse that is outlined in Section~\ref{sec:review} into a unifying definition:
\emph{\textbf{Loop engineering} is the practice of designing automated control structures that repeatedly invoke coding agents, triggered on a schedule or by events, with each agent loop bounded by a machine-checkable stop condition. A well-engineered loop persists state across agent loops, verifies results independently of the implementation agents, bounds inference costs, restricts what an unattended agent loop may change without approval, and defines when humans should intervene.}

\begin{table}
\centering
\footnotesize
\caption{Four layers of guiding AI coding agents~\cite{osmani2026loop, huashu2026playbook, galster2026harness}. Each layer subsumes the previous one; the harness engineering layer refers to the outer harness.}
\label{tab:layers}
\begin{tabularx}{\columnwidth}{llX}
\toprule
\textbf{Layer} & \textbf{Unit of Concern} & \textbf{Example Artifacts} \\
\midrule
Prompt eng. & One instruction & Prompts, prompt templates \\
Context eng. & One model call & \texttt{AGENTS.md}, \texttt{CLAUDE.md}, rules \\
Harness eng. & One agent loop & Skills, subagents, hooks, MCP config. \\
Loop eng. & Repeated agent loops & Schedules, stop conditions, state files, run logs, budgets, verifier agents \\
\bottomrule
\end{tabularx}
\end{table}

\subsection{Adjacent Academic Work}
\label{sec:background-related}

\looseness=-1 In a preprint, \citeauthor{macedo2026loops} defined a \enq{loop specification} as a reusable artifact made of a trigger, a goal, a verification step, a stopping rule, and a memory. They hand-coded 50 entries of a public loop catalog along five categories: trigger, goal type, verification level, architecture, and terminal states~\cite{macedo2026loops}. Such a study based on a loop catalog can show how published loop designs are composed, but not how many projects run certain loops. Therefore, our mining study complements such studies by estimating how many engineered open-source software projects run loops (Section~\ref{sec:agenda-mining}).
Related to loop engineering, \citeauthor{geng2026asynchronous} studied \emph{asynchronous} software engineering agents that operate without a human in the loop and reported that a design combining dependency-aware plans, isolated workspaces, and test-based verification improved task accuracy~\cite{geng2026asynchronous}.
Benchmarks such as SWE-bench evaluate agents on individual software engineering tasks~\cite{jimenez2024swebench}.
We, however, study the developer practice of designing recurring control structures in the software projects they maintain.
Existing autonomy taxonomies grade how much a developer delegates to an agent: \citeauthor{feng2025levels} define five autonomy levels based on the user's role, from operator to observer~\cite{feng2025levels}.
\citeauthor{cihon2025measuring} measure autonomy from the code that orchestrates the model, which, in our terminology, would be the loop~\cite{cihon2025measuring}.
A randomized controlled trial found experienced open-source developers measurably \emph{slowed down} by early-2025 tools, despite believing the opposite~\cite{becker2025measuring}. Whether recurring loops change such results is an open question.

\section{Exploratory Literature Review}
\label{sec:review}

\looseness=-1 Because the term \emph{loop engineering} was coined by practitioners only weeks before we wrote this article, the relevant knowledge resides almost exclusively in gray literature, a situation for which established review guidelines recommend including such sources~\cite{garousi2019guidelines}.
Having followed the discourse on social media since the seed posts of June 2026, the last author compiled a first wave of seven sources published between 7~June and 15~July 2026: Osmani's blog post~\cite{osmani2026loop} and one-month retrospective~\cite{osmani2026monthon}, Orosz's newsletter investigation~\cite{orosz2026newsletter} and its 136-comment discussion thread~\cite{orosz2026post}, Lindenberg's newsletter article~\cite{lindenberg2026loops}, a self-published playbook~\cite{huashu2026playbook}, and a snapshot of Greyling's community repository~\cite{greyling2026repo}.
On 18~July, we added three sources that the collected material referenced: Huntley's post introducing the Ralph technique~\cite{huntley2025ralph}, Greyling's essay accompanying his repository~\cite{greyling2026essay}, and the podcast interview in which Stripe engineer Steve Kaliski describes the company's internal agent system~\cite{kaliski2026podcast}. On 25~July, after consolidating the observations below, we added one further source, Osmani's post on agent autonomy and a related slide deck~\cite{osmani2026autonomy}.
We included sources that treat loop engineering substantively (define the term, describe building blocks or patterns, or report experiences) and excluded passing mentions, reactions, and reposts.


\looseness=-1 In parallel, on 15~July 2026, we searched arXiv, DBLP, and Semantic Scholar for uses of the term \emph{loop engineering} and for adjacent academic work (Section~\ref{sec:background-related}). An LLM-based agentic tool (Claude Code with Fable~5) assisted in this search as well as in locating the sources that the collected material referenced.
The tool further assisted in extracting candidate quotes, which the last author then checked against the sources. Fable~5 was trained on data up to January 2026, five months before loop engineering was coined~\cite{anthropic2026fable5}, so Claude Code worked from the supplied and retrieved sources, not from prior knowledge of the term or its discourse. We disclose this in line with the community guidelines for empirical studies involving LLMs~\cite{baltes2026guidelines}.
From the collected sources, we extracted proposed definitions, building blocks, claimed benefits, risks, and explicitly skeptical positions, which the last author consolidated into the six observations below (O1--O6), reviewed by all authors.


\paragraph{O1: The definitions converge on handing off the next prompt.}
\looseness=-1 Across sources, the definitional core of loop engineering is stable: The developer stops deciding what the agent does next and designs the system that decides instead~\cite{osmani2026loop, lindenberg2026loops, huashu2026playbook}.
The four-part taxonomy Osmani reported (Section~\ref{sec:background-emergence}) operationalizes this core as a sequence of hand-offs~\cite{osmani2026monthon}.
In Orosz's thread, one commenter argued that an agent started on a schedule or by an event is \enq{still largely a workflow}: It becomes a loop only when an explicit feedback signal feeds one iteration's outcome into the next~\cite{orosz2026post}.

\paragraph{O2: The novelty of the practice is contested.}
\looseness=-1 Many reactions in the collected sources reject the framing of loop engineering as novel: Loops are \enq{a renamed cron job}, and the term repackages event-driven architecture with, as one commenter put it, \enq{a fuzzy worker}~\cite{orosz2026post}.
The same commenter argued that the novelty of loop engineering comes from having a non-deterministic worker, requiring engineered \enq{guardrails around it}~\cite{orosz2026post}.
\looseness=-1 Some developers suspect the AI labs' interests behind a practice that consumes many tokens, a suspicion Orosz discusses under the heading \enq{tokenmaxxing}~\cite{orosz2026newsletter}.
Orosz's own investigation, drawing on roughly 210 practitioner replies, concluded that most concrete examples of loops fall into two familiar buckets, i.e., cron jobs and event-based triggers, which are now executed by more capable workers~\cite{orosz2026newsletter, orosz2026post}.
Max Kanat-Alexander expects agentic coding tools to absorb loop engineering, since anything that developers must know about a tool \enq{would eventually become baked into its default workflow}~\cite{orosz2026newsletter}.
To Orosz, with \texttt{/loop} and \texttt{/goal} now built-in, loop engineering already seemed \enq{as good as obsolete} outside AI infrastructure~\cite{orosz2026newsletter}.
However, by our definition, designing loops using built-in mechanisms is still loop engineering.
In summary, the discourse has not yet concluded whether loop engineering is a new discipline, a transitional technique, or a marketing label.

\paragraph{O3: The proposed building blocks are consistent across sources.}
\looseness=-1 Despite the definitional dispute, the sources largely agree on what a well-engineered loop contains: a trigger (a schedule or an event) and a stop condition that decides when an agent loop is done, durable state outside the context window, encoded project knowledge (skills), isolation for parallel work (worktrees), an independent verifier separated from the implementer, connectors to external systems, budgets with a documented way to pause a running loop (a \enq{kill switch}~\cite{greyling2026repo}), binding constraints on what an unattended agent loop may change without human approval, and defined escalation points to humans~\cite{osmani2026loop, lindenberg2026loops, greyling2026repo, huashu2026playbook}.
However, the practitioner sources are not independent. Greyling's community repository and HuaShu's playbook explicitly build on Osmani's posts, and Lindenberg cites the same seed posts, so much of the agreement traces back to one origin.

\looseness=-1 The most emphasized building block of loop engineering is the \enq{maker/checker} separation. Osmani calls the existence of a verifier the reason one \enq{can walk away} (while the loop is running)~\cite{osmani2026loop}, Greyling's verifier skill rejects by default~\cite{greyling2026repo}, and Lindenberg warns that a loop \enq{satisfies the gate you wrote}, not the more abstract goal behind it~\cite{lindenberg2026loops}.

The sources also recommend staged adoption. Greyling's readiness levels start report-only (L1) and reach unattended operation (L3) only once the verifier has proven reliable~\cite{greyling2026repo}.
Osmani's later post revises the eight stages of developer evolution that Steve Yegge had published in January 2026 (from near-zero AI use to building one's own orchestrator)~\cite{yegge2026gastown}.
Osmani recommends choosing an autonomy level for each task and distinguishes \emph{agency}, how far one agent proceeds without human intervention, from \emph{orchestration}, how many agents run at once and who coordinates them~\cite{osmani2026autonomy}.
Goal-driven autonomy (L3) is the highest agency level in Osmani's classification. The agent iterates until a measurable stopping condition holds. L4 and L5 describe parallel delegation and managed-by-exception orchestration. A team should raise autonomy only as far as it can verify the result at acceptable cost~\cite{osmani2026autonomy}.

\paragraph{O4: Claimed benefits are large but are based on self-reports and anecdotes.}
\looseness=-1 The reported use cases are plausible and concrete: flaky-test stabilization, nightly end-to-end test repair, and incremental migrations run from cron jobs~\cite{orosz2026newsletter}.
The boldest claims, however, are self-reported.
For example, in a podcast interview, Stripe engineer Steve Kaliski describes the company's internal \enq{Minions} system as producing about 1{,}300 agent-written pull requests per week~\cite{kaliski2026podcast}, and Osmani reports that Mozilla used a loop to produce 423 security fixes for Firefox in one month~\cite{osmani2026monthon}.
No published study or dataset we know of supports these numbers, and HuaShu's playbook states that other numbers circulating in the discourse \enq{are mostly secondhand summaries and should be treated as rough reference}~\cite{huashu2026playbook}.

\paragraph{O5: The discourse comprehensively documents failure modes.}
\looseness=-1 Reported problems include high token usage (one report describes a `CI sweeper' loop using roughly eight million tokens in 48 hours), infinite fix loops, verifiers that approve without real checks (\enq{verifier theater}), notification fatigue, and drift between a loop's committed instructions and its state~\cite{greyling2026repo}, as well as plain disappointment from developers who tried loops and found they did not help~\cite{orosz2026newsletter}.
Two risks were described under the labels \emph{comprehension debt}, i.e., the gap between what exists in the repository and what the developer understands grows with loop velocity, and \emph{cognitive surrender}, i.e., using loops to avoid thinking rather than to move faster on understood work~\cite{osmani2026loop, greyling2026essay}.
Comprehension debt is the individual counterpart of what \citeauthor{storey2026debt} calls \emph{cognitive debt}, a team's eroding shared understanding of the system it changes. She places cognitive debt beside \emph{technical debt} and \emph{intent debt}, the unrecorded goals and rationale that developers and coding agents need to change a system safely~\cite{storey2026debt}.
Commenters in Orosz's thread observe that nightly agent loops can open more pull requests than teams can review, so the volume itself defeats the \enq{reviewed by devs} guardrail that the loop depends on. Review fatigue turns the human gate into a \enq{rubber stamp} while \enq{the pipeline still reports green}. Another commenter argues that loops remain in use only when someone is already paid to review their output~\cite{orosz2026post}.

\paragraph{O6: Systematic evidence on the value claims is missing.}
\looseness=-1 The discourse agrees on a design vocabulary but makes conflicting claims about the benefits and costs of loop engineering. The collected sources provide no systematic evidence to evaluate these claims.
Together, O1--O6 answer \textbf{RQ1}: The discourse defines loop engineering as designing automated control structures that repeatedly invoke coding agents, agrees on its building blocks, and disputes its novelty and value.

\section{Empirically Studying Loop Engineering}
\label{sec:agenda}

\subsection{Mining Loop Engineering Traces (RQ2)}
\label{sec:agenda-mining}

\looseness=-1 Table~\ref{tab:traceability} summarizes the loop mechanisms we searched for, together with related repository artifacts, and the number of repositories where our heuristics matched artifacts for each mechanism. The result columns count repositories with detected matches (Detected), confirmed matches (TP), and confirmed matches and evidence of loop execution (TP+L). For each matched repository, we checked workflow run histories and agent outputs for evidence that an autonomous agent loop had been executed. Evidence of loop execution is per repository, not per artifact.

\begin{table*}
\centering
\scriptsize
\caption{Loop engineering mechanisms and their traceability from GitHub data. The first eight rows are the building blocks named in the practitioner sources (Section~\ref{sec:review}); the last two are traces of a loop's operation. `Detected': number of repositories with matches (\emph{n}~=~36,487); `TP' (true positives): repositories with a match that we confirmed. `TP+L': TP plus evidence that an autonomous agent loop ran. We did not validate provenance; local worktrees are not visible in GitHub data.}
\label{tab:traceability}
\begin{tabular}{>{\raggedright\arraybackslash}p{2.3cm}>{\raggedright\arraybackslash}p{3.8cm}>{\raggedright\arraybackslash}p{4.1cm}>{\raggedright\arraybackslash}p{4.1cm}@{\hspace{4pt}}r@{\hspace{4pt}}r@{\hspace{4pt}}r@{}}
\toprule
\textbf{Mechanism} & \textbf{Role in the Loop} & \textbf{Example Repository Artifacts} & \textbf{Traceability from GitHub Data} & \textbf{Detected} & \textbf{TP} & \textbf{TP+L} \\
\midrule
Triggering \& scheduling & Starts agent loops on a cadence or event & \texttt{.github/workflows/*.yml} with \texttt{cron} schedules invoking agent CLIs & CI schedules visible; tool-native schedulers (\texttt{/loop}) and local cron jobs leave no trace & 253 & 242 & 215 \\
Goal \& stop conditions & Specifies when an agent loop is complete, in a form a machine can check & Goal definitions in skills, commands, or workflow scripts & Visible only when encoded in committed prompts or scripts & 0 & 0 & 0 \\
State \& memory & Persists findings and progress across agent loops & \texttt{STATE.md}, \texttt{LOOP.md}, plan/progress/log files & Files plus commit history revealing cadence, authorship, and append-only updates & 2 & 0 & 0 \\
Skills \& intent & Encodes durable project knowledge read in every agent loop & \texttt{.claude/skills/}, \texttt{SKILL.md} & Presence detectable; runtime usage invisible~\cite{galster2026harness} & 1 & 1 & 0 \\
Parallel isolation & Lets concurrent agents work without collisions & Git worktrees (local); branch naming conventions & Worktrees are not pushed; agent branches indicate general agent use, not loop isolation & --- & --- & --- \\
Verification (\enq{maker/checker}) & Checks the stop condition independently of the implementer; can reject & Verifier subagents (\texttt{.claude/agents/}), CI gates, protected branches & Definitions and CI results visible; agent-side verdicts invisible unless logged & 0 & 0 & 0 \\
Human oversight \& escalation & Requires human approval for defined actions; escalates on demand & Pull-request reviews, \texttt{CODEOWNERS}, escalation issues, gate rules in loop docs & Review traces and documented gates visible; interactive supervision invisible & 15 & 13 & 10 \\
Budgets \& pause controls & Caps spend per agent loop; restricts what an unattended agent loop may change & \texttt{loop-budget.md}, \texttt{loop-constraints.md}, pause labels & Declarations visible when committed; enforcement invisible & 0 & 0 & 0 \\
\midrule
Outcomes \& provenance & Identifies what the loop produced & Bot-authored commits/PRs, \texttt{Co-Authored-By} trailers & Attribution conventions inconsistent across tools and teams~\cite{robbes2026agentic, siddiq2026security}; attribution to loop executions not assessed & --- & --- & --- \\
Cost & Tokens and money actually consumed & Run logs with token estimates & Rarely committed; observed logs hold template values, not measurements~\cite{greyling2026repo} & 0 & 0 & 0 \\
\bottomrule
\end{tabular}
\end{table*}

\paragraph{Detection.}
\looseness=-1 Our mining study builds on a repository sample ($n=36,710$) from an AI coding tool configuration dataset we built earlier. These repositories were classified with an LLM as engineered software projects, out of a sampling frame of actively maintained repositories in ten programming languages ($n=40,585$)~\cite{galster2026dataset, baltes2026dataset}.

\looseness=-1 On 19~August 2026, we searched every repository in this sample for committed loop artifacts with deterministic file- and content-based \emph{heuristics}, and tested the commit history of candidate context and state files for three behavioral \emph{signals}, defined below.
We searched each repository at its current head commit and dated each detected artifact by its introducing commit. We found that 65 repositories (0.18\%) were no longer reachable on GitHub. We further excluded 158 duplicate repositories (0.43\%, see Section~\ref{sec:threats}), resulting in a sample of 36,487 repositories.
We used a \emph{trigger heuristic} to identify scheduled or event-triggered invocations of eight coding tools in committed files: five tools that we considered when building the above-mentioned AI configuration dataset (Claude Code, GitHub Copilot, Cursor, OpenAI Codex, and Gemini CLI),  plus OpenCode, Hermes, and OpenClaw. We added these three open-source tools because they provide built-in scheduling or headless run-until-done commands that our heuristics can recognize in committed files~\cite{greyling2026repo}.

\looseness=-1 Loop building blocks can also occur as ordinary outer-harness configuration. For example, finding a skill describing how to run tests does not tell us whether that skill is part of a loop. Our heuristics therefore required evidence linking artifacts to loops: scheduled or event-triggered agent invocations, skills or verifier subagents referenced by loop artifacts, and dated run-log entries or combinations of loop-related content in state files.
For GitHub Actions, the trigger heuristic excluded workflows started only by a human mention, label, or manual dispatch (\texttt{workflow\_dispatch}), which we treated as interactive agent use. It retained workflows declaring \texttt{schedule}, \texttt{workflow\_run}, or \texttt{repository\_dispatch}.
The supplementary material provides the complete heuristics and code used to apply them~\cite{lulla2026supplement}.

\looseness=-1 Dedicated loop artifacts are rare. We therefore complemented our search for committed loop artifacts with behavioral signals from the commit histories of context files (e.g., \texttt{AGENTS.md}, \texttt{CLAUDE.md}) and candidate state files (e.g., \texttt{STATE.md}, \texttt{LOOP.md}, plan and progress files). Loops that commit updates to these files can leave patterns in their commit histories. In each file's latest 15 commits (or all commits if fewer), we looked for three signals: (1)~\emph{cadence}, i.e., a regular inter-commit interval between ten minutes and seven days; (2)~\emph{authorship}, i.e., a strict majority of commits attributed to one of the eight coding tools; and (3)~\emph{append-only}, i.e., edits that add dated run-log entries with few deletions.
Because maintenance bots produce similar commit patterns and humans also edit context and state files~\cite{mohsenimofidi2026context}, we excluded 12 maintenance bots and classified a repository as loop-adopting only if \emph{authorship} co-occurred with one of the other two signals on a detected state or budget artifact.
We also treated co-occurring signals in a context file as a reason to inspect the repository, even without a detected loop artifact.

For our commit-history analysis, we selected repositories with artifact matches or candidate state files. From our AI coding tool configuration dataset, we also selected repositories with context files that had at least five commits, including at least one attributed to a coding tool. This yielded 792 repositories. In 637 of them, we analyzed 1,334 commit histories of context or candidate state files with at least five commits.

\paragraph{Validation and inspection.}
\looseness=-1 We drafted the heuristics and thresholds from Greyling's repository and checked tool-invocation patterns and the trigger heuristic against a 100-file GitHub code-search sample. The sample contained no state or budget files on which to test the state and budget heuristics. We inspected every matched repository to estimate precision, but did not construct a labeled set of repositories without loops to estimate false-positive rates. The supplementary material explains the thresholds and reports per-file signal measurements~\cite{lulla2026supplement}.

\looseness=-1 In the full repository sample, our heuristics produced 341 artifact matches across 255 repositories. With guidance from Claude Code, the last author reviewed these matches to check whether each file contained the claimed loop mechanism, confirming 325 matches.

Authorship and cadence co-occurred in a context file in one additional repository. We inspected all 256 repositories for evidence of autonomous agent loop execution. Using a codebook included in our supplementary material~\cite{lulla2026supplement}, Claude Code with Fable 5 judged whether each repository runs an autonomous agent loop (\emph{yes}, \emph{no}, or \emph{unclear}); with Opus 5, it then tried to refute these labels and changed disputed \emph{yes} or \emph{no} labels to \emph{unclear}. The last author then approved or revised these labels. We disclose this LLM-as-annotator design following the community guidelines~\cite{baltes2026guidelines}.

For adjudication, we archived GitHub Actions run histories, job and step records, and evidence of outputs such as commits and issue comments on 19 and 20~August 2026. The last author checked the model-assigned labels for autonomous agent loop execution against these archived records and retained 196 of the 256 labels.

GitHub marked some workflow runs as successful even though the agent step was skipped or ended without doing the requested work. The last author checked the agent step's duration and output to determine whether the agent had actually run.
For example, \href{https://github.com/AI-Hypercomputer/xpk/blob/a097189c80dc197006818acbc78cfced1ee0d8ce/.github/workflows/gemini-scheduled-triage.yml}{\texttt{AI-Hypercomputer/xpk}}'s Gemini triage logged 6,290 workflow runs, but the search for issues to triage returned none at the head commit we examined, and none of the workflow runs that we checked reached the agent step.

\looseness=-1 The archived GitHub Actions histories also showed Copilot review runs without corresponding committed workflow files. GitHub offers an account-level setting that requests a Copilot review automatically and leaves no trace in the repository~\cite{github2026copilotreview}. In 95 of the 159 repositories for which we measured review timing, Copilot reviews clustered on individual authors' pull requests, most arriving within minutes of one being opened. That pattern suggests the account setting rather than a repository-wide rule. We did not count this pattern as adoption by the repository because an individual contributor's account settings can generate reviews across projects. We confirmed a loop only on committed configuration or where Copilot's review volume approached the repository's pull request volume.

\paragraph{Loop execution.}
\looseness=-1 The last author's review confirmed autonomous agent loops in 217 of the 256 repositories that we inspected, 0.59\% of all analyzed repositories. 
We excluded two repositories from \enq{TP+L} because they had no confirmed artifact match (Table~\ref{tab:traceability}).
We found no autonomous loop in 33 of the inspected repositories. For the remaining six repositories, we could not determine whether an autonomous agent loop had run. The precision of our repository detection was 0.87 (217/250), excluding these six repositories.

The most common tool in the confirmed loops was Claude Code (189 of 217), followed by OpenAI Codex (11), OpenCode (6), Gemini CLI (5), Cursor (4), and Copilot (2).

We grouped the confirmed loops by the trigger that accounted for most of their archived workflow runs: 21 ran on a schedule (e.g., \href{https://github.com/dimagi/commcare-android/blob/8ddc9359cda0cf5b76ccad9d8771d6483538b6a8/.github/workflows/pr-comment-handler.lock.yml}{\texttt{dimagi/commcare-android}}, with 276 scheduled workflow runs in the archived GitHub Actions history), 180 ran on repository events (mostly automatic pull request reviews), 15 on both, and one ran until its GitHub Actions workflow broke. In 12 of the 217 confirmed loops, every trigger workflow that we detected was marked as manually disabled. We counted those loops and the one with the broken workflow as confirmed because they demonstrably ran.

\paragraph{Detected mechanisms.}
\looseness=-1 Table~\ref{tab:traceability} reports how many of the 36,487 analyzed repositories had artifact matches for each mechanism. The trigger heuristic matched 253 repositories (35 scheduled only, 205 event-triggered only, 13 both). The last author's file review confirmed agent-invoking trigger definitions in 242 of these repositories. All but one of the 323 trigger matches were GitHub Actions workflows; the exception was a committed shell script.

\looseness=-1 For state and memory (Table~\ref{tab:traceability}), our searches of committed files and their histories found no loop state files or updates on the default branches that we scanned. Our artifact heuristics matched only two state files, one skill referenced by a loop artifact, no stop condition, no budget file, no verifier subagent referenced by a loop artifact, and no cost log with measured values. We rejected both state files on inspection as homonyms of unrelated \enq{loop} vocabulary. The one skill match was genuine: an issue-triage skill written for a loop. Its repository's committed agent workflows required a human request for each run, so we did not count them as autonomous loops. In 96 repositories, we found a skill with a name or content indicating a loop, and in 18 a subagent with a name or content indicating a verifier, but no loop workflow or loop document in those repositories referred to these skills and subagents.

\looseness=-1 Claude Code used workflow definitions, prompts, and archived GitHub Actions run histories to annotate the repositories. The annotations recorded a goal or stop condition in 55 of the 256 inspected repositories, including committed prompts that told the agent when to stop. Table~\ref{tab:traceability} reports no goal artifact because we matched a fixed list of stop-condition phrases against the workflow or document that started a loop and the skills anchored to it, and none of these phrases occurred in those files. The annotations also recorded state persisted across agent loops in 35 of those repositories, a spending or turn limit in 101, a verifier in 76, and a cost record in 10. We also checked for the autonomy levels described in the practitioner literature (Section~\ref{sec:review}), but no inspected repository declared an autonomy level for its loops.

\looseness=-1 For parallel isolation (Table~\ref{tab:traceability}), GitHub data includes branch names but not the local Git worktrees used to separate concurrent agents' work. We found 3,507 repositories with two or more branches under agent name prefixes (e.g., \texttt{claude/}). These branches indicate coding-agent use, but do not establish whether a loop isolated parallel work~\cite{robbes2026agentic}.

\looseness=-1 For outcomes and provenance (Table~\ref{tab:traceability}), the commit histories showed coding-tool contributions to context files. Among the context files that we analyzed, 384 files in 217 repositories had a majority of their latest 15 commits (or all commits if fewer) attributed to coding tools. Of these repositories, 30 also had a detected trigger, state, or budget artifact. Only 25 of the 217 repositories had confirmed autonomous loops. These attributions did not establish that agent loops produced the commits to the context files; we did not link the commits to specific loop executions.

\looseness=-1 Together, these results answer \textbf{RQ2}: Committed loop artifacts primarily documented triggers. Although the artifact heuristics found no goals, budgets, or verifiers, Claude Code recorded goals in 55 of the 256 inspected repositories, budgets in 101, and verifiers in 76. However, neither the searches of committed files nor the commit-history checks provided evidence of the persistent loop state files suggested by practitioners.

\subsection{Loop Goals and Feedback (RQ3)}
\label{sec:agenda-experiment}

\looseness=-1 To answer \textbf{RQ3}, we propose a controlled experiment focusing on individual agent loops (Figure~\ref{fig:experiment-design}). The experiment varies three \emph{factors}: loop goals, static analysis feedback, and a decision policy. The loop goal factor has two \emph{levels}: no goal and a goal to resolve the provided issue. Static analysis feedback is independently disabled or enabled. The decision policy is either `accept-all' or a simulated developer. The eight combinations of factor levels define the \emph{treatment conditions}. We compare outcomes between treatment conditions that differ in only one factor, keeping the other two factors fixed. All eight conditions use the same harness and model versions and the same time and token limits. We will publish the prompts and tool parameters for every treatment condition. The design follows established guidelines for experimentation~\cite{wohlin2012experimentation} and empirical studies involving LLMs~\cite{baltes2026guidelines}.

\begin{figure*}[t]
\centering
\scalebox{0.95}{%
\begin{tikzpicture}[
  remember picture,
  x=1cm,y=1cm,
  font=\sffamily\fontsize{8.5}{10}\selectfont,
  box/.style={draw=black!70,rounded corners=1.5pt,fill=black!3,
    align=left,inner sep=2mm},
  left box/.style={box,text width=2.45cm},
  right box/.style={box,text width=2.65cm},
  treatment/.style={box,dashed,fill=white,font=\sffamily\bfseries},
  flow/.style={-{Latex[length=1.8mm]},line width=0.55pt},
  parameter/.style={flow,dashed,black!70},
  dependency/.style={line width=0.55pt,black!35},
  edge label/.style={font=\sffamily\scriptsize,inner sep=1.5pt,outer sep=0pt}
]
\def\rqFrameGap{4mm}
\def\rqItemAnchor#1{\tikz[remember picture,overlay,baseline=0pt]{\coordinate (#1) at (0,0.5ex);}}
\def\rqItemLabelAnchor#1{\makebox[0pt][r]{\rqItemAnchor{#1}\phantom{\labelenumi}\hspace{\labelsep}}}

\node[left box] (codingtool) at (5.55,-2.0) {
  \textbf{\rqFixed\ Agentic tool}
  \begin{itemize}[leftmargin=1em,labelsep=0.4em,nosep,topsep=1mm]
    \item Read/write/edit
    \item Use tools
  \end{itemize}
};
\node[right box,anchor=north] (checks) at ($(codingtool.north)+(3.85,0)$) {
  \textbf{\rqFixed\ Checks}
  \begin{itemize}[leftmargin=1em,labelsep=0.4em,nosep,topsep=1mm]
    \item Regression tests
  \end{itemize}
};
\draw[flow] (codingtool.east |- checks.west) -- node[edge label,above,xshift=-0.5mm]
  {build} (checks.west);
\coordinate (reportinput) at ($(codingtool.north)+(0.55,0)$);
\coordinate (reportlevel) at ([yshift=4mm]codingtool.north);
\draw[flow] (checks.north) -- (checks.north |- reportlevel)
  -- node[edge label,above] {feedback}
  (reportinput |- reportlevel) -- (reportinput);

\node[left box]
  (intervention) at (5.55,-4.55) {
  \textbf{Responses}\\
  Upon input request
};
\draw[flow] ($(codingtool.south)+(-0.25,0)$) --
  node[edge label,left] {input request}
  ($(intervention.north)+(-0.25,0)$);
\draw[flow] ($(intervention.north)+(0.25,0)$) --
  node[edge label,right] {feedback}
  ($(codingtool.south)+(0.25,0)$);
\node[right box,anchor=south] (finish) at (checks.center |- intervention.south) {
  \textbf{\rqFixed\ Stop conditions}
  \begin{itemize}[leftmargin=1em,labelsep=0.4em,nosep,topsep=1mm]
    \item Tool finishes\\(no loop goal)
    \item Goal achieved
    \item Time/token limit
  \end{itemize}
};
\coordinate (endturn) at ($(codingtool.east)!0.5!(checks.west)$);
\draw[flow] ([yshift=-2mm]codingtool.east)
  -| (endturn |- finish.west) -- (finish.west);
\coordinate (patchlevel) at (0,-2.8);
\coordinate (patchbend) at ($(finish.east |- patchlevel)+(0.25,0)$);
\begin{scope}[on background layer]
\node[box,fill=white,fit=(codingtool)(checks)(intervention)(finish),
  inner sep=5mm,inner ysep=4.5mm,yshift=2.5mm] (runframe) {};
\end{scope}
\node[treatment,text width=3.1cm,align=left,
  font=\sffamily\fontsize{8.5}{10}\selectfont,anchor=south west]
  (goals) at ($(runframe.north west)+(0,\rqFrameGap)$) {
  \textbf{\rqVaried\ Goals}
  \begin{enumerate}[leftmargin=*,labelsep=0.35em,nosep,topsep=1mm]
    \item No loop goal
    \item Resolve the issue
    \item \ldots{}and fix new static\rqItemAnchor{goalwarning-text}\\
      analysis warnings
  \end{enumerate}
};
\node[treatment,text width=2.8cm,align=left,
  font=\sffamily\fontsize{8.5}{10}\selectfont,anchor=south east]
  (feedback) at ($(runframe.north east)+(0,\rqFrameGap)$) {
  \textbf{\rqVaried\ Feedback}
  \begin{enumerate}[leftmargin=*,labelsep=0.35em,nosep,topsep=1mm]
    \item Test results
    \item \rqItemLabelAnchor{analysisfeedback-text}\ldots{}and static\\
      analysis warnings
  \end{enumerate}
};
\draw[parameter] ($(goals.south -| codingtool.north)+(-0.55,0)$) --
  ($(codingtool.north)+(-0.55,0)$);
\coordinate (feedbackinput) at ([xshift=7mm]checks.north);
\draw[parameter] (goals.south -| feedbackinput) -- (feedbackinput);
\node[box,text width=3.5cm,anchor=east] (task) at
  ($(runframe.west |- codingtool.center)+(-\rqFrameGap,0)$) {
  \textbf{\rqFixed\ Task inputs}
  \begin{itemize}[leftmargin=1em,labelsep=0.4em,nosep,topsep=1mm]
    \item Issue description\\(initial prompt)
    \item Repository\\(at reference commit)
    \item Regression tests
  \end{itemize}
};
\coordinate (goalwarning) at ([xshift=1mm]goalwarning-text);
\coordinate (analysisfeedback) at ([xshift=-0.6mm]analysisfeedback-text);
\coordinate (analysislink) at ($(goals.east)!0.5!(feedback.west)$);
\draw[dependency] (goalwarning) -- (analysisfeedback);
\node[dependency,edge label,fill=white] at (analysislink |- goalwarning)
  {\rqIcon{\faLink}};
\foreach \rqEndpoint in {analysisfeedback,goalwarning} {
  \fill[dependency] (\rqEndpoint) circle[radius=1.25pt];
}
\draw[flow] (task.east) -- (codingtool.west);
\node[treatment,text width=3.5cm,align=left,font=\sffamily,anchor=south east]
  (decisions) at (task.east |- runframe.south) {
  \textbf{\rqVaried\ Decisions}
  {\sffamily\fontsize{8.5}{10}\selectfont
  \begin{enumerate}[leftmargin=*,labelsep=0.35em,nosep,topsep=1mm]
    \item Accept-all policy
    \item Simulated developer
  \end{enumerate}}
};
\draw[parameter] (decisions.east |- intervention.west) -- (intervention.west);
\node[box,text width=4.95cm,anchor=west] (evaluation) at
  ($(runframe.east |- patchbend)+(\rqFrameGap,0)$) {
  \textbf{\rqFixed\ Final evaluation}
  \begin{enumerate}[leftmargin=1.3em,labelsep=0.35em,
    itemsep=1mm,topsep=1mm,parsep=0pt]
    \item Apply patch to clean checkout
    \item Run tests
      \begin{itemize}[leftmargin=0.8em,labelsep=0.35em,nosep,topsep=0pt]
        \item Benchmark tests: \mbox{fail\,$\to$\,pass?}
        \item Regression tests: \mbox{pass\,$\to$\,pass?}
      \end{itemize}
    \item Report static analysis results\\(\#warnings before/after patch)
    \item Report resource usage\\(max memory, wallclock/CPU time)
  \end{enumerate}
};
\node[edge label,above,fill=white] at
  (runframe.east |- patchbend) {patch};
\draw[flow] (finish.east) -- (patchbend |- finish.east)
  -- (patchbend) -- (evaluation.west);
\end{tikzpicture}
}
\caption{Experimental design for RQ3: Goals, tool feedback, and decision policy vary~(\rqVaried{}). The issue description is the initial task prompt in every agent loop. When setting a loop goal, we explicitly require resolving the issue. With static analysis feedback, we extend the goal to require fixing new static analysis warnings~(\rqIcon{\faLink}). Each decision policy is tested with and without a loop goal and with and without static analysis feedback, giving eight treatment conditions. Benchmark tests are used only for final scoring, but agents may create and run additional tests in every treatment condition. Task inputs, harness and model versions, time/token limits, and final evaluation remain fixed~(\rqFixed{}).}
\Description{Three dashed boxes show loop goals, tool feedback, and decision policy. The loop goal is either absent or explicitly set to resolve the issue. Static analysis feedback is independently disabled or enabled during agent loops. When a loop goal is set and static analysis feedback is enabled, we extend the goal to require fixing warnings introduced since the reference commit. The two decision policies are the accept-all policy and the simulated developer. These three binary factors give eight treatment conditions. The issue description supplies the initial task prompt. Each agent loop starts with a repository at its reference commit and regression tests. Agents may create and run additional tests in every treatment condition. The agentic coding tool reads, writes, and edits files and can run build scripts. Test output and, when enabled, static analysis output are provided as tool feedback. User-input requests are answered under the selected decision policy. The stop conditions are tool completion for agent loops without a loop goal, goal achievement for agent loops with a goal, and time or token limits for all agent loops. Final evaluation applies the patch to a clean checkout and runs benchmark tests and regression tests. It reports static analysis warnings before and after each patch in every treatment condition and measures maximum memory, wallclock time, and CPU time during test execution.}
\label{fig:experiment-design}
\end{figure*}

\looseness=-1 We will select tasks from SWE-bench~\cite{jimenez2024swebench} or similar benchmarks. Each task starts from the benchmark's \emph{reference commit}. Tests that pass at that commit serve as \emph{regression tests}. The issue-resolution tests supplied by the benchmark serve as \emph{benchmark tests} for final scoring. Before collecting developer interactions, we will check that benchmark tests fail at the reference commit and pass with the benchmark's reference patch, while regression tests remain passing. We will also manually compare the issue description, test assertions, and reference patch to identify requirements without tests and assertions that impose unstated requirements. Issues with unexpected test results or identified mismatches will be excluded~\cite{openai2026verified,openai2026audit}. These checks may still miss test defects. The supplementary material for the planned journal extension will specify the selected benchmark release and the version or commit of the code used to score patches.

\looseness=-1 While the agent loop is running, regression tests are available in the repository checkout. We apply the benchmark's test changes only when scoring the final patch produced by the agentic coding tool~\cite{jimenez2024swebench}.
\emph{Benchmark contamination} remains possible if benchmark tests or solutions were included in the LLM's training data~\cite{xu2024contamination}.
Such contamination can inflate benchmark scores~\cite{openai2026verified}.
The agentic coding tool uses the same LLM version in all experimental conditions, so any training-data contamination is shared. This makes contamination less concerning for comparisons between conditions than for absolute scores, although it could still affect conditions differently.

\looseness=-1 Agents may create and run additional tests in every treatment condition, even without an explicit instruction to do so. We will record all agent actions in interaction traces to determine retrospectively whether and when agents created additional tests.

\looseness=-1 We use the issue title and body as the initial task prompt in every treatment condition~\cite{jimenez2024swebench}.
All experimental conditions use the same system prompt and tool definitions. In conditions with a loop goal, we explicitly set a goal using \texttt{/goal} or an equivalent command. The default goal is to resolve the provided issue (\emph{Goals} in Figure~\ref{fig:experiment-design}). With static analysis feedback enabled, we extend the goal to also require fixing new warnings. We identify new warnings by comparing static analysis results with those at the reference commit. Without an explicit loop goal, we assess the effect of static analysis feedback while keeping the issue description as the initial task prompt. With an explicit goal to resolve the issue, we assess the combined effect of static analysis feedback and the requirement to fix new warnings.

\looseness=-1 For the repositories from which the selected benchmark tasks were derived, we will check existing build scripts and modify them as needed to run the regression tests and, when enabled, static analysis. We will add a build script if none exists. When the agentic coding tool runs these scripts, we provide their output as feedback (\emph{Feedback} in Figure~\ref{fig:experiment-design}).

\looseness=-1 The \emph{simulated developer} responds to user-input requests from the agentic coding tool (\emph{Decisions} in Figure~\ref{fig:experiment-design}). It uses an LLM guided by an agent skill or context file~\cite{galster2026configuring} derived from recorded developer interactions to provide responses similar to those of human developers. As a simple baseline, we use an accept-all policy to approve proposals automatically. For mutually exclusive or open-ended choices, we ask the agentic coding tool to recommend an option and automatically accept its recommendation.

\looseness=-1 To derive such an agent skill or context file, we will record developers' responses to user-input requests while they work with an agentic coding tool on selected benchmark tasks. Participants will be recruited from industry contacts, open-source contributors, and experienced students. Published interaction logs of developer-agent sessions may provide an additional source for deriving the agent skill or context file. We will validate the simulated developer on held-out sessions from different tasks. For each user-input request in these sessions, we will compare the simulated developer's response with the recorded human response.

\looseness=-1 We will report the proportion of evaluated user-input requests for which the simulated developer's response agrees with the recorded human response. For fixed-choice requests, agreement means selecting the same option. We will compare free-text responses using an LLM-as-judge approach. If agreement falls below a minimum agreement threshold specified before starting the evaluation, we will omit the simulated developer and retain the accept-all baseline. We may then compare accept-all with a baseline that selects uniformly at random among permitted option combinations.
If we use the simulated developer, we will repeat the experiment with a second LLM for that role. We will check whether the differences between experimental conditions change in size or direction for each outcome measure.

\looseness=-1 The experiment uses three \emph{Stop conditions} shown in Figure~\ref{fig:experiment-design}. For agent loops without a loop goal, the agentic coding tool's final response stops the loop. For agent loops with a loop goal, the loop stops when the goal is achieved. All agent loops also stop if they reach the specified time or token limit.

\looseness=-1 In the \emph{Final evaluation} (Figure~\ref{fig:experiment-design}), we save the changes made by the agentic coding tool as a patch when a stop condition is met. We then apply this patch to a clean checkout of the task's repository at the reference commit. To assess task success, we run the benchmark and regression tests. Task success requires all of these tests to pass~\cite{jimenez2024swebench}.
An agent loop counts as unsuccessful if it did not result in a patch or if the patch fails any benchmark test or regression test.

Besides functional correctness, we will assess the code quality of the generated patch.
To this end, we will count static analysis warnings before and after each patch, including added and removed warnings.
We will use static analysis tools appropriate to each language, with comparable rule sets and severity thresholds across languages.
Tool versions and configurations will remain fixed across experimental conditions.

Another non-functional property that we want to measure is resource usage.
During the final test execution, we measure wallclock time, maximum memory use, and CPU time using identical hardware and execution environments. We also record the duration of each agent loop and input and output token consumption for the agentic coding tool and the simulated developer.
We will then compare test-execution costs across experimental conditions per benchmark task.
For this comparison, we retain only tasks for which all conditions being compared produced patches that pass the tests.

\section{Discussion}
\label{sec:discussion}

We discuss the implications of our findings for software engineering researchers, practitioners, and educators.

\paragraph{Implications for researchers.}
\looseness=-1 Our gray literature review identified recommendations for loop goals, independent verification, and budgets. However, the reviewed sources did not provide systematic evidence of their effects. Researchers evaluating the effectiveness and cost of loop engineering could compare agent loops with and without explicit goals, independent verification, or budgets, measuring task success and resource usage. Our proposed experiment tests explicit goals, static analysis feedback, and decision policies for user-input requests on individual issue-resolution tasks.

\looseness=-1 Researchers studying the adoption of autonomous agent loops need to verify agent execution because some successful GitHub Actions runs in our dataset skipped their agent step. A follow-up adoption study could survey maintainers of repositories in our mining sample about their use of agent loops. Researchers could then invite maintainers who report that they use agent loops to share interaction traces and scheduler logs, including for loops whose configuration is not committed.

\looseness=-1 These execution records could also support longitudinal studies of agent loops used for pull request review and maintenance, the main tasks identified in our mining study. Linking successive agent loops to their starting commits, verification results, and resulting changes could reveal repeated reviews of unchanged code or reversals of earlier changes. Recording the time, token consumption, and developer interventions associated with repeated reviews and reversals could quantify the cost of this rework across successive agent loops.

\paragraph{Implications for practitioners.}
\looseness=-1 Teams planning to implement agent loops for maintenance could test verification approaches from our gray literature review. For example, they could compare verification by the agent making changes with verification by a second agent. The comparison could show differences in the proportion of patches that pass tests and receive developer approval, as well as in resource usage and token consumption.

\looseness=-1 Even when a second agent verifies changes, teams need to decide when agents should ask developers to clarify requirements. In our gray literature review, Ng emphasized developers' product and user knowledge as a \enq{context advantage}~\cite{ng2026loops}.
Teams could therefore configure escalation policies that require agents to ask developers about unresolved product requirements even when automated checks pass.

\paragraph{Implications for educators.}
\looseness=-1 Educators in universities and professional developer training could ask learners to set a loop goal and budget for a maintenance task. Learners could then inspect execution records and resulting changes to check whether the goal was achieved within the specified budget.

\paragraph{Beyond individual loops.}
\looseness=-1 On 17~July 2026, Peter Steinberger asked whether the discussion had shifted from loops to graphs yet~\cite{steinberger2026graphs}. Within hours, Hamel Husain published \enq{Loop Engineering Is Dead. Enter Graph Engineering}~\cite{husain2026graphs}. Lindenberg describes graph engineering as connecting loops in which specialized agents perform different tasks and share state~\cite{lindenberg2026graphs}.
\looseness=-1 To study graph engineering, experiments could keep individual loop designs fixed while varying task assignment and information sharing between loops. Outcomes could include task success, resource usage, and token consumption.

\section{Threats to Validity}
\label{sec:threats}

\paragraph{Construct validity.}
\looseness=-1 Construct validity concerns whether our measures capture the intended properties~\cite{ralph2018construct}.
Definitions in our gray literature review differed in how they distinguished loop engineering from other agent-invoking automation. For our mining study, our working definition and codebook required autonomous agent execution, excluding workflows that needed a person to start every agent loop. We included loops that had executed before being disabled, so the adoption rate does not measure how many loops were active on the collection date.

For the \emph{authorship} signal, which we defined to identify files predominantly updated by coding tools, we inferred coding-tool contributions from names, email addresses, and commit trailers. We excluded 12 maintenance bots and counted \texttt{github-actions[bot]} commits only when their messages referred to a workflow run and the repository had a detected agent workflow. We did not match these run references to specific GitHub Actions runs, so we could misattribute commits from unrelated workflows to a coding tool. We mitigated the effect of commit-attribution errors on our estimate of autonomous agent loop adoption by requiring the last author to confirm loop execution using archived GitHub Actions run histories and agent outputs.

\paragraph{Internal validity.}
\looseness=-1 Internal validity concerns the extent to which we can rule out alternative explanations for our results~\cite{brewer2000validity}.
For the gray literature review, source selection and interpretation by the last author could have favored prominent practitioners or favorable experiences. We included skeptical positions and failure reports, the last author checked extracted quotes, and all authors reviewed the six observations. Several posts reused the same seed sources, so agreement among them was not independent corroboration.

Our mining study's artifact heuristics (Section~\ref{sec:agenda-mining}) might miss loops with unrecognized file conventions or tool invocations. We checked trigger and tool-invocation patterns against a 100-file sample, and the last author reviewed all 341 artifact matches, rejecting 16 false positives. These checks cannot establish how many loops the searches missed. We did not inspect a sample of repositories without matches or qualifying behavioral signals, so the reported adoption rate may underestimate autonomous agent loop adoption.

The capture limits of our data collection pipeline omitted files in 26 repositories and truncated large files in 795 repositories, potentially hiding loop artifacts. The affected repositories are identified in the supplementary material. Our search of committed files covered only default branches. For example, \href{https://github.com/dimagi/commcare-android/blob/8ddc9359cda0cf5b76ccad9d8771d6483538b6a8/.github/workflows/pr-comment-handler.lock.yml}{\texttt{dimagi/commcare-android}}'s workflow stored memory on the \texttt{memory/pr-comment-handler} branch, outside that search. Our commit-history checks required at least five commits per candidate file, so they could not detect patterns in files with shorter histories.

The last author checked the LLM-assigned loop labels against archived run histories and agent outputs, but read the model's judgments before making his own decisions. Those judgments could have influenced him, so the label-retention rate does not measure agreement between independent annotators. In six inspected repositories, he could not determine whether an autonomous agent loop had run. We excluded these repositories when calculating precision.

\paragraph{External validity.}
\looseness=-1 External validity concerns the extent to which our results generalize to other contexts~\cite{baltes2022sampling}.
We estimated the adoption rate of autonomous agent loops in a repository sample from the AI coding tool configuration dataset~\cite{galster2026dataset}.
This sample was drawn in March 2026 from actively maintained projects in ten programming languages. Searching its current head commits in August updated the repository data, but did not add projects outside the original sample. The results do not establish prevalence across all open-source software or the internal company systems described in our gray literature review. The artifact heuristics recognized only the eight coding tools listed in Section~\ref{sec:agenda-mining}, excluding loops implemented through other tools or direct LLM API calls. We excluded repositories with identical head commits to reduce duplication, but forks with different head commits may still contain copies of the same agent workflow files.

\paragraph{Data availability.}
\looseness=-1 Our supplementary material is openly available on Zenodo~\cite{lulla2026supplement}.
It contains the mining scripts, rule catalog, result data, annotation protocol and codebook, adjudication record, per-repository evidence files, and archived GitHub Actions run histories. It also includes a glossary, a snapshot of the MIT-licensed loop engineering reference repository, and a source list linking to the archived academic publications, practitioner posts, and tool documentation.

\section{Conclusion}
\label{sec:conclusion}

\looseness=-1 Our gray literature (\textbf{RQ1}) review found broad agreement on the building blocks of loop engineering but no systematic evidence for its claimed benefits.
Our mining study (\textbf{RQ2}) extended \citeauthor{galster2026harness}'s analysis of harness configuration~\cite{galster2026harness} to autonomous agent loops, which we confirmed in 217 open-source repositories.
Most of the identified loops reviewed pull requests or ran scheduled maintenance. We found no loop state files on the default branches that we scanned.

\looseness=-1 Our proposed experiment (\textbf{RQ3}) will test how explicit loop goals, static analysis feedback, and decision policies for user-input requests affect task success, code quality, resource usage, and token consumption. It is a first step toward closing the evidence gap identified in our gray literature review. In the journal extension, we will run this experiment and repeat the gray literature review and mining study to examine changes in definitions and adoption.

\looseness=-1 Future studies of loop engineering should consider execution traces that connect goals, state updates, verification, and the changes produced by agents. We call on software engineering researchers to measure how autonomous agent loops affect issue resolution, code quality, development costs, and developers' understanding of their software.


\balance 
\bibliographystyle{ACM-Reference-Format}
\bibliography{literature}

@misc{white2023promptpatterns,
  author        = {Jules White and Quchen Fu and Sam Hays and Michael Sandborn and Carlos Olea and Henry Gilbert and Ashraf Elnashar and Jesse Spencer-Smith and Douglas C. Schmidt},
  title         = {A Prompt Pattern Catalog to Enhance Prompt Engineering with {ChatGPT}},
  year          = {2023},
  eprint        = {2302.11382},
  archivePrefix = {arXiv},
  primaryClass  = {cs.SE},
  doi           = {10.48550/arXiv.2302.11382},
  url           = {https://arxiv.org/abs/2302.11382}
}

@misc{schulhoff2024promptreport,
  author        = {Sander Schulhoff and Michael Ilie and Nishant Balepur and Konstantine Kahadze and Amanda Liu and Chenglei Si and Yinheng Li and Aayush Gupta and others},
  title         = {The Prompt Report: A Systematic Survey of Prompt Engineering Techniques},
  year          = {2024},
  eprint        = {2406.06608},
  archivePrefix = {arXiv},
  doi           = {10.48550/arXiv.2406.06608},
  url           = {https://arxiv.org/abs/2406.06608}
}

@inproceedings{tafreshipour2025prompting,
  author       = {Mahan Tafreshipour and Aaron Imani and Eric Huang and Eduardo Santana de Almeida and Thomas Zimmermann and Iftekhar Ahmed},
  title        = {Prompting in the Wild: An Empirical Study of Prompt Evolution in Software Repositories},
  booktitle    = {22nd {IEEE/ACM} International Conference on Mining Software Repositories, {MSR} 2025},
  pages        = {686--698},
  publisher    = {{IEEE}},
  address      = {Ottawa, ON, Canada},
  year         = {2025},
  doi          = {10.1109/MSR66628.2025.00106}
}

@inproceedings{villamizar2025prompts,
  author       = {Hugo Villamizar and
                  Jannik Fischbach and
                  Alexander Korn and
                  Andreas Vogelsang and
                  Daniel M{\'{e}}ndez},
  title        = {Prompts as Software Engineering Artifacts: {A} Research Agenda and
                  Preliminary Findings},
  booktitle    = {Product-Focused Software Process Improvement - 26th International
                  Conference, {PROFES} 2025, Salerno, Italy, December 1-3, 2025, Proceedings},
  series       = {Lecture Notes in Computer Science},
  pages        = {470--478},
  publisher    = {Springer},
  address      = {Cham},
  year         = {2025},
  doi          = {10.1007/978-3-032-12089-2_32}
}

@misc{mei2025context,
  author        = {Lingrui Mei and
                   Jiayu Yao and
                   Yuyao Ge and
                   Yiwei Wang and
                   Baolong Bi and
                   Yujun Cai and
                   Jiazhi Liu and
                   Mingyu Li and
                   Zhong{-}Zhi Li and
                   Duzhen Zhang and
                   Chenlin Zhou and
                   Jiayi Mao and
                   Tianze Xia and
                   Jiafeng Guo and
                   Shenghua Liu},
  title         = {A Survey of Context Engineering for Large Language Models},
  year          = {2025},
  eprint        = {2507.13334},
  archivePrefix = {arXiv},
  primaryClass  = {cs.CL},
  doi           = {10.48550/arXiv.2507.13334},
  url           = {https://arxiv.org/abs/2507.13334}
}

@misc{hua2025context20,
  author        = {Qishuo Hua and Lyumanshan Ye and Dayuan Fu and Yang Xiao and Xiaojie Cai and Yunze Wu and Jifan Lin and Junfei Wang and Pengfei Liu},
  title         = {Context Engineering 2.0: The Context of Context Engineering},
  year          = {2025},
  eprint        = {2510.26493},
  archivePrefix = {arXiv},
  doi           = {10.48550/arXiv.2510.26493},
  url           = {https://arxiv.org/abs/2510.26493}
}

@misc{mohsenimofidi2026context,
  author        = {Seyedmoein Mohsenimofidi and Matthias Galster and Christoph Treude and Sebastian Baltes},
  title         = {Context Engineering for {AI} Agents in Open-Source Software},
  year          = {2025},
  eprint        = {2510.21413},
  archivePrefix = {arXiv},
  primaryClass  = {cs.SE},
  doi           = {10.48550/arXiv.2510.21413},
  url           = {https://arxiv.org/abs/2510.21413},
  note          = {To appear at the 23rd IEEE/ACM International Conference on Mining Software Repositories (MSR 2026), Rio de Janeiro, Brazil}
}

@misc{lulla2026agentsmd,
  author        = {Jai Lal Lulla and
                   Seyedmoein Mohsenimofidi and
                   Matthias Galster and
                   Jie M. Zhang and
                   Sebastian Baltes and
                   Christoph Treude},
  title         = {On the Impact of {AGENTS.md} Files on the Efficiency of {AI} Coding Agents},
  year          = {2026},
  eprint        = {2601.20404},
  archivePrefix = {arXiv},
  primaryClass  = {cs.SE},
  doi           = {10.48550/arXiv.2601.20404},
  url           = {https://arxiv.org/abs/2601.20404},
  note          = {To appear at the 1st Journal Ahead Workshop ({JAWs}@{ICSE} 2026)}
}

@misc{macedo2026loops,
  author        = {Sandeco Macedo},
  title         = {Stop Hand-Holding Your Coding Agent: Engineering the Loops that Replace Step-by-Step Prompting},
  year          = {2026},
  eprint        = {2607.00038},
  archivePrefix = {arXiv},
  primaryClass  = {cs.SE},
  doi           = {10.48550/arXiv.2607.00038},
  url           = {https://arxiv.org/abs/2607.00038},
  note          = {Preprint, submitted 28 June 2026}
}

@misc{chatlatanagulchai2025agentreadmes,
  author        = {Worawalan Chatlatanagulchai and
                   Hao Li and
                   Yutaro Kashiwa and
                   Brittany Reid and
                   Kundjanasith Thonglek and
                   Pattara Leelaprute and
                   Arnon Rungsawang and
                   Bundit Manaskasemsak and
                   Bram Adams and
                   Ahmed E. Hassan and
                   Hajimu Iida},
  title         = {Agent {READMEs}: An Empirical Study of Context Files for Agentic Coding},
  year          = {2025},
  eprint        = {2511.12884},
  archivePrefix = {arXiv},
  primaryClass  = {cs.SE},
  doi           = {10.48550/arXiv.2511.12884},
  url           = {https://arxiv.org/abs/2511.12884}
}

@inproceedings{galster2026configuring,
  author       = {Galster, Matthias and Mohsenimofidi, Seyedmoein and Lulla, Jai Lal and Abubakar, Muhammad Auwal and Treude, Christoph and Baltes, Sebastian},
  title        = {Configuring Agentic {AI} Coding Tools: An Exploratory Study},
  booktitle    = {Proceedings of the 3rd ACM International Conference on AI-Powered Software (AIware '26)},
  pages        = {11--20},
  year         = {2026},
  publisher    = {ACM},
  address      = {Montreal, QC, Canada},
  isbn         = {979-8-4007-2601-9},
  doi          = {10.1145/3805760.3814887},
  url          = {https://doi.org/10.1145/3805760.3814887}
}

@misc{galster2026harness,
  author        = {Galster, Matthias and Mohsenimofidi, Seyedmoein and Lulla, Jai Lal and Abubakar, Muhammad Auwal and Treude, Christoph and Baltes, Sebastian},
  title         = {Harness Engineering for Agentic {AI} Coding Tools: An Exploratory Study},
  year          = {2026},
  eprint        = {2602.14690},
  archivePrefix = {arXiv},
  primaryClass  = {cs.SE},
  doi           = {10.48550/arXiv.2602.14690},
  url           = {https://arxiv.org/abs/2602.14690},
  note          = {Extended version of the AIware '26 paper ``Configuring Agentic AI Coding Tools: An Exploratory Study''}
}

@misc{ning2026codeharness,
  author        = {Xuying Ning and Katherine Tieu and Dongqi Fu and Tianxin Wei and Zihao Li and Yuanchen Bei and Jiaru Zou and Mengting Ai and others},
  title         = {Code as Agent Harness: Toward Executable, Verifiable, and Stateful Agent Systems},
  year          = {2026},
  eprint        = {2605.18747},
  archivePrefix = {arXiv},
  primaryClass  = {cs.CL},
  doi           = {10.48550/arXiv.2605.18747},
  url           = {https://arxiv.org/abs/2605.18747}
}

@misc{geng2026asynchronous,
  author        = {Jiayi Geng and Graham Neubig},
  title         = {Effective Strategies for Asynchronous Software Engineering Agents},
  year          = {2026},
  eprint        = {2603.21489},
  archivePrefix = {arXiv},
  doi           = {10.48550/arXiv.2603.21489},
  url           = {https://arxiv.org/abs/2603.21489}
}

@inproceedings{jimenez2024swebench,
  author       = {Carlos E. Jimenez and
                  John Yang and
                  Alexander Wettig and
                  Shunyu Yao and
                  Kexin Pei and
                  Ofir Press and
                  Karthik R. Narasimhan},
  title        = {{SWE-bench}: Can Language Models Resolve Real-world Github Issues?},
  booktitle    = {The Twelfth International Conference on Learning Representations,
                  {ICLR} 2024},
  publisher    = {OpenReview.net},
  address      = {Vienna, Austria},
  year         = {2024},
  url          = {https://openreview.net/forum?id=VTF8yNQM66},
  numpages     = {51},
  note         = {OpenReview proceedings are unpaginated; numpages from the camera-ready PDF}
}

@misc{openai2026verified,
  author       = {{OpenAI}},
  title        = {Why {SWE-bench Verified} no longer measures frontier coding capabilities},
  year         = {2026},
  howpublished = {Blog post},
  url          = {https://openai.com/index/why-we-no-longer-evaluate-swe-bench-verified/},
  note         = {Published 2026-02-23, accessed 2026-07-20}
}

@misc{openai2026audit,
  author       = {{OpenAI}},
  title        = {Separating signal from noise in coding evaluations},
  year         = {2026},
  howpublished = {Blog post},
  url          = {https://openai.com/index/separating-signal-from-noise-coding-evaluations/},
  note         = {Published 2026-07-08, accessed 2026-07-20}
}

@misc{xu2024contamination,
  author       = {Cheng Xu and Shuhao Guan and Derek Greene and M-Tahar Kechadi},
  title        = {Benchmark Data Contamination of {Large Language Models}: A Survey},
  year         = {2024},
  eprint       = {2406.04244},
  archivePrefix = {arXiv},
  doi          = {10.48550/arXiv.2406.04244},
  url          = {https://arxiv.org/abs/2406.04244}
}

@misc{becker2025measuring,
  author        = {Joel Becker and Nate Rush and Elizabeth Barnes and David Rein},
  title         = {Measuring the Impact of Early-2025 {AI} on Experienced Open-Source Developer Productivity},
  year          = {2025},
  eprint        = {2507.09089},
  archivePrefix = {arXiv},
  doi           = {10.48550/arXiv.2507.09089},
  url           = {https://arxiv.org/abs/2507.09089}
}

@misc{gloaguen2026evaluating,
  author        = {Thibaud Gloaguen and Niels M{\"u}ndler and Mark M{\"u}ller and Veselin Raychev and Martin Vechev},
  title         = {Evaluating {AGENTS.md}: Are Repository-Level Context Files Helpful for Coding Agents?},
  year          = {2026},
  eprint        = {2602.11988},
  archivePrefix = {arXiv},
  primaryClass  = {cs.SE},
  doi           = {10.48550/arXiv.2602.11988},
  url           = {https://arxiv.org/abs/2602.11988}
}

@misc{robbes2026agentic,
  author        = {Romain Robbes and
                   Th{\'{e}}o Matricon and
                   Thomas Degueule and
                   Andr{\'{e}} C. Hora and
                   Stefano Zacchiroli},
  title         = {Agentic Much? Adoption of Coding Agents on GitHub},
  year          = {2026},
  eprint        = {2601.18341},
  archivePrefix = {arXiv},
  primaryClass  = {cs.SE},
  doi           = {10.48550/arXiv.2601.18341},
  url           = {https://arxiv.org/abs/2601.18341}
}

@misc{siddiq2026security,
  author        = {Mohammed Latif Siddiq and Xinye Zhao and Vinicius Carvalho Lopes and Beatrice Casey and Joanna C. S. Santos},
  title         = {Security in the Age of {AI} Teammates: An Empirical Study of Agentic Pull Requests on GitHub},
  year          = {2026},
  eprint        = {2601.00477},
  archivePrefix = {arXiv},
  doi           = {10.48550/arXiv.2601.00477},
  url           = {https://arxiv.org/abs/2601.00477}
}

@misc{feng2025levels,
  author        = {K. J. Kevin Feng and David W. McDonald and Amy X. Zhang},
  title         = {Levels of Autonomy for {AI} Agents},
  year          = {2025},
  eprint        = {2506.12469},
  archivePrefix = {arXiv},
  doi           = {10.48550/arXiv.2506.12469},
  url           = {https://arxiv.org/abs/2506.12469}
}

@misc{cihon2025measuring,
  author        = {Peter Cihon and Merlin Stein and Gagan Bansal and Sam Manning and Kevin Xu},
  title         = {Measuring {AI} Agent Autonomy: Towards a Scalable Approach with Code Inspection},
  year          = {2025},
  eprint        = {2502.15212},
  archivePrefix = {arXiv},
  doi           = {10.48550/arXiv.2502.15212},
  url           = {https://arxiv.org/abs/2502.15212}
}

@misc{sapkota2025vibe,
  author        = {Ranjan Sapkota and Konstantinos I. Roumeliotis and Manoj Karkee},
  title         = {Vibe Coding vs. Agentic Coding: Fundamentals and Practical Implications of Agentic {AI}},
  year          = {2025},
  eprint        = {2505.19443},
  archivePrefix = {arXiv},
  doi           = {10.48550/arXiv.2505.19443},
  url           = {https://arxiv.org/abs/2505.19443}
}

@misc{ge2025vibecoding,
  author        = {Yuyao Ge and Lingrui Mei and Zenghao Duan and Tianhao Li and Yujia Zheng and Yiwei Wang and others},
  title         = {A Survey of Vibe Coding with Large Language Models},
  year          = {2025},
  eprint        = {2510.12399},
  archivePrefix = {arXiv},
  doi           = {10.48550/arXiv.2510.12399},
  url           = {https://arxiv.org/abs/2510.12399}
}

@incollection{brewer2000validity,
  author       = {Marilynn B. Brewer},
  title        = {Research Design and Issues of Validity},
  booktitle    = {Handbook of Research Methods in Social and Personality Psychology},
  editor       = {Harry T. Reis and Charles M. Judd},
  pages        = {3--16},
  publisher    = {Cambridge University Press},
  address      = {Cambridge, UK},
  year         = {2000},
  isbn         = {978-0-521-55903-4},
  url          = {https://assets.cambridge.org/97805215/59034/excerpt/9780521559034_excerpt.pdf}
}

@article{baltes2022sampling,
  author       = {Sebastian Baltes and Paul Ralph},
  title        = {Sampling in Software Engineering Research: A Critical Review and Guidelines},
  journal      = {Empirical Software Engineering},
  volume       = {27},
  number       = {4},
  articleno    = {94},
  numpages     = {31},
  year         = {2022},
  doi          = {10.1007/s10664-021-10072-8},
  url          = {https://doi.org/10.1007/s10664-021-10072-8}
}

@inproceedings{ralph2018construct,
  author       = {Paul Ralph and Ewan Tempero},
  title        = {Construct Validity in Software Engineering Research and Software Metrics},
  booktitle    = {Proceedings of the 22nd International Conference on Evaluation and Assessment in Software Engineering 2018},
  series       = {EASE '18},
  pages        = {13--23},
  publisher    = {Association for Computing Machinery},
  address      = {New York, NY, USA},
  year         = {2018},
  doi          = {10.1145/3210459.3210461},
  url          = {https://doi.org/10.1145/3210459.3210461}
}

@article{garousi2019guidelines,
  author       = {Vahid Garousi and Michael Felderer and Mika V. M{\"a}ntyl{\"a}},
  title        = {Guidelines for including grey literature and conducting multivocal
                  literature reviews in software engineering},
  journal      = {Information and Software Technology},
  volume       = {106},
  pages        = {101--121},
  year         = {2019},
  doi          = {10.1016/j.infsof.2018.09.006}
}

@misc{baltes2026guidelines,
  author        = {Sebastian Baltes and Florian Angermeir and Chetan Arora and Marvin Mu{\~n}oz Bar{\'o}n and Chunyang Chen and Lukas B{\"o}hme and Fabio Calefato and Neil Ernst and Davide Falessi and Brian Fitzgerald and Davide Fucci and Junda He and Christoph Treude and Marcos Kalinowski and Stefano Lambiase and Daniel Russo and Mircea Lungu and Cristina Martinez Montes and Lutz Prechelt and Paul Ralph and Rijnard van Tonder and Stefan Wagner},
  title         = {Guidelines for Empirical Studies in Software Engineering involving Large Language Models},
  year          = {2026},
  eprint        = {2508.15503},
  archivePrefix = {arXiv},
  doi           = {10.48550/arXiv.2508.15503},
  url           = {https://arxiv.org/abs/2508.15503},
  note          = {Accepted for publication in Empirical Software Engineering}
}

@book{wohlin2012experimentation,
  author       = {Claes Wohlin and Per Runeson and Martin H{\"o}st and Magnus C. Ohlsson and Bj{\"o}rn Regnell and Anders Wessl{\'e}n},
  title        = {Experimentation in Software Engineering},
  publisher    = {Springer},
  address      = {Berlin, Heidelberg},
  year         = {2012},
  doi          = {10.1007/978-3-642-29044-2},
  isbn         = {978-3-642-29043-5}
}

@misc{agentsmd2026,
  author       = {{Agentic AI Foundation}},
  title        = {{AGENTS.md}},
  year         = {2026},
  howpublished = {Website},
  url          = {https://agents.md/},
  note         = {Stewarded by the Agentic AI Foundation under the Linux Foundation, accessed 2026-07-18}
}

@misc{huntley2025ralph,
  author       = {Geoffrey Huntley},
  title        = {Ralph {Wiggum} as a ``software engineer''},
  year         = {2025},
  howpublished = {Blog post},
  url          = {https://ghuntley.com/ralph/},
  note         = {Published 2025-07-14, accessed 2026-07-18}
}

@misc{kaliski2026podcast,
  author       = {Claire Vo},
  title        = {How {Stripe} built ``minions''---{AI} coding agents that ship 1,300 {PRs} weekly from {Slack} reactions},
  year         = {2026},
  howpublished = {Podcast episode, How I AI},
  url          = {https://podcasts.apple.com/us/podcast/how-stripe-built-minions-ai-coding-agents-that-ship/id1809663079?i=1000757255000},
  note         = {Interview with Steve Kaliski (Stripe). Published 2026-03-25, accessed 2026-07-18}
}

@misc{cherny2026post,
  author       = {Boris Cherny},
  title        = {I'm {Boris} and {I} created {Claude Code}.},
  year         = {2026},
  howpublished = {X post},
  url          = {https://x.com/bcherny/status/2007179832300581177},
  note         = {Published 2026-01-02, accessed 2026-07-18}
}

@misc{cherny2026unplugged,
  author       = {{WorkOS}},
  title        = {{Boris Cherny}: {Claude Code} \& the Future of Engineering},
  year         = {2026},
  howpublished = {Video interview, Acquired Unplugged, WorkOS YouTube channel},
  url          = {https://www.youtube.com/watch?v=RkQQ7WEor7w},
  note         = {Interview with Boris Cherny (Anthropic). Published 2026-06-02, accessed 2026-08-21}
}

@misc{boeckeler2026harness,
  author       = {Birgitta B{\"o}ckeler},
  title        = {Harness Engineering for Coding Agent Users},
  year         = {2026},
  howpublished = {Blog post},
  url          = {https://martinfowler.com/articles/harness-engineering.html},
  note         = {Published 2026-04-02, accessed 2026-07-17}
}

@misc{osmani2026loop,
  author       = {Addy Osmani},
  title        = {Loop Engineering},
  year         = {2026},
  howpublished = {Blog post},
  url          = {https://addyosmani.com/blog/loop-engineering/},
  note         = {Published 2026-06-07, accessed 2026-07-15}
}

@misc{osmani2026monthon,
  author       = {Addy Osmani},
  title        = {Loop engineering, one month on},
  year         = {2026},
  howpublished = {LinkedIn post},
  url          = {https://www.linkedin.com/posts/addyosmani_loop-engineering-a-month-on-ugcPost-7483004445094506496-R62H/},
  note         = {Published 2026-07-15, accessed 2026-07-15}
}

@misc{osmani2026autonomy,
  author       = {Addy Osmani},
  title        = {{AI} coding? {The} autonomy you grant an agent is the level you can cheaply verify},
  year         = {2026},
  howpublished = {LinkedIn post with slide deck},
  url          = {https://www.linkedin.com/posts/addyosmani_agentic-autonomy-levels-ugcPost-7486670405957505024-ZjLa/},
  note         = {Published 2026-07-25, accessed 2026-07-26}
}

@misc{yegge2026gastown,
  author       = {Steve Yegge},
  title        = {Welcome to Gas Town},
  year         = {2026},
  howpublished = {Blog post on Medium},
  url          = {https://steve-yegge.medium.com/welcome-to-gas-town-4f25ee16dd04},
  note         = {Published 2026-01-01, accessed 2026-07-27}
}

@misc{steinberger2026post,
  author       = {Peter Steinberger},
  title        = {Here's your monthly reminder that you shouldn't be prompting coding agents anymore. {You} should be designing loops that prompt your agents.},
  year         = {2026},
  howpublished = {X post},
  url          = {https://x.com/steipete/status/2063697162748260627},
  note         = {Published 2026-06-07, accessed 2026-07-15}
}

@misc{orosz2026newsletter,
  author       = {Gergely Orosz},
  title        = {What is ``loop engineering?''},
  year         = {2026},
  howpublished = {The Pragmatic Engineer newsletter},
  url          = {https://newsletter.pragmaticengineer.com/p/what-is-loop-engineering},
  note         = {Published 2026-07-14, accessed 2026-07-18}
}

@misc{orosz2026post,
  author       = {Gergely Orosz},
  title        = {What is ``loop engineering'' to you, anyway?},
  year         = {2026},
  howpublished = {LinkedIn post},
  url          = {https://www.linkedin.com/posts/gergelyorosz_what-is-loop-engineering-to-you-anyway-share-7482371640924737536-tmm8/},
  note         = {Published 2026-07-13, accessed 2026-07-15. The archived capture shows no absolute date; the post ID decodes to 2026-07-13 09:53 UTC. The same decoding reproduces the recorded dates of both Osmani LinkedIn posts exactly, and the post solicits the replies the 14 July newsletter reports on}
}

@misc{lindenberg2026loops,
  author       = {Andr{\'e} Lindenberg},
  title        = {From Prompts to Loops: The Next Step of Agent Engineering},
  year         = {2026},
  howpublished = {LinkedIn newsletter article, Artificial Engineering \#61},
  url          = {https://www.linkedin.com/pulse/rom-prompts-loops-next-step-agent-engineering-andr%C3%A9-lindenberg-wnqre/},
  note         = {Published 2026-06-28, accessed 2026-07-15}
}

@misc{lindenberg2026graphs,
  author       = {Andr{\'e} Lindenberg},
  title        = {From Loops to Graphs},
  year         = {2026},
  howpublished = {LinkedIn newsletter article, Artificial Engineering \#65},
  url          = {https://www.linkedin.com/pulse/from-loops-graphs-andr%C3%A9-lindenberg-m8jae},
  note         = {Published 2026-07-25, accessed 2026-07-28}
}

@misc{ng2024agentic,
  author       = {Andrew Ng},
  title        = {Robots Talk Back, {AI} Security Risks, Political Deepfakes, and more},
  year         = {2024},
  howpublished = {The Batch, Issue 241, DeepLearning.{AI}},
  url          = {https://www.deeplearning.ai/the-batch/issue-241},
  note         = {Published 2024-03-20, accessed 2026-07-28}
}

@misc{ng2026loops,
  author       = {Andrew Ng},
  title        = {Loop Engineering: Three Key Loops for Building Great Software},
  year         = {2026},
  howpublished = {The Batch, DeepLearning.{AI}},
  url          = {https://www.deeplearning.ai/the-batch/three-key-loops-for-building-great-software},
  note         = {Published 2026-06-26, accessed 2026-07-28}
}

@misc{steinberger2026graphs,
  author       = {Peter Steinberger},
  title        = {Are we still talking loops or did we shift to graphs yet?},
  year         = {2026},
  howpublished = {X post},
  url          = {https://x.com/steipete/status/2078277297791189132},
  note         = {Published 2026-07-17, accessed 2026-07-28}
}

@misc{husain2026graphs,
  author       = {Hamel Husain},
  title        = {Loop Engineering Is Dead. Enter Graph Engineering},
  year         = {2026},
  howpublished = {X article},
  url          = {https://x.com/HamelHusain/article/2078346425621237935},
  note         = {Published 2026-07-18, accessed 2026-08-21}
}

@misc{greyling2026repo,
  author       = {Cobus Greyling},
  title        = {loop-engineering: Design systems that prompt your agents},
  year         = {2026},
  howpublished = {GitHub repository},
  url          = {https://github.com/cobusgreyling/loop-engineering},
  note         = {Accessed 2026-07-15}
}

@misc{greyling2026essay,
  author       = {Cobus Greyling},
  title        = {Loop Engineering},
  year         = {2026},
  howpublished = {Substack essay},
  url          = {https://cobusgreyling.substack.com/p/loop-engineering},
  note         = {Published 2026-06-09, accessed 2026-07-18}
}

@misc{huashu2026playbook,
  author       = {{HuaShu}},
  title        = {Loop Engineering: The {Anthropic} Playbook for Designing Systems That Prompt Your Agents},
  year         = {2026},
  howpublished = {Self-published working note (Orange Books, v260615)},
  url          = {https://huasheng.ai/orange-books},
  note         = {Independent conference-style reformatting of the open guide ``Loop Engineering: Stop Asking Me What It Is''; not peer-reviewed. Accessed 2026-07-15}
}

@misc{baltes2026dataset,
  author       = {Sebastian Baltes and
                  Seyedmoein Mohsenimofidi and
                  Levi B\"{o}hme and
                  Jai Lal Lulla and
                  Muhammad Auwal Abubakar and
                  Christoph Treude and
                  Matthias Galster},
  title        = {{A Dataset of Agentic AI Coding Tool Configurations}},
  month        = apr,
  year         = 2026,
  publisher    = {Zenodo},
  doi          = {10.5281/zenodo.19375880},
  url          = {https://doi.org/10.5281/zenodo.19375880}
}

@misc{galster2026dataset,
  author        = {Matthias Galster and
                  Seyedmoein Mohsenimofidi and
                  Levi B\"{o}hme and
                  Jai Lal Lulla and
                  Muhammad Auwal Abubakar and
                  Christoph Treude and
                  Sebastian Baltes},
  title         = {{A Dataset of Agentic AI Coding Tool Configurations}},
  year          = {2026},
  eprint        = {2605.08435},
  archivePrefix = {arXiv},
  primaryClass  = {cs.SE},
  doi           = {10.48550/arXiv.2605.08435},
  url           = {https://arxiv.org/abs/2605.08435},
  note          = {Dataset archived on Zenodo, DOI 10.5281/zenodo.19375880}
}

@dataset{lulla2026supplement,
  author       = {Jai Lal Lulla and
                  Vahram Nersesyan and
                  Seyedmoein Mohsenimofidi and
                  Christoph Treude and
                  Sebastian Baltes},
  title        = {{Loop Engineering: Building Blocks, Adoption, and Impact (Supplementary Material)}},
  month        = jul,
  year         = {2026},
  organization = {Zenodo},
  doi          = {10.5281/zenodo.21540692},
  url          = {https://doi.org/10.5281/zenodo.21540692}
}

@article{storey2026debt,
  author       = {Margaret-Anne Storey},
  title        = {From Technical Debt to Cognitive and Intent Debt},
  journal      = {Queue},
  volume       = {24},
  number       = {2},
  numpages     = {15},
  month        = mar,
  year         = {2026},
  doi          = {10.1145/3807966},
  url          = {https://doi.org/10.1145/3807966}
}

@misc{anthropic2026fable5,
  author       = {{Anthropic}},
  title        = {{Claude Fable 5}},
  year         = {2026},
  howpublished = {Claude Platform Docs, model overview},
  url          = {https://platform.claude.com/docs/en/models/fable-5/overview},
  note         = {Accessed 2026-09-15}
}

@misc{github2026copilotreview,
  author       = {{GitHub}},
  title        = {{Configuring code review by GitHub Copilot}},
  year         = {2026},
  howpublished = {GitHub Docs},
  url          = {https://docs.github.com/en/copilot/how-tos/copilot-on-github/set-up-copilot/configure-code-review},
  note         = {Accessed 2026-09-15}
}

\end{document}